\documentclass[pre,aps,onecolumn,showpacs,floatfix,10pt,superscriptaddress]{revtex4}

\usepackage{amssymb}
\usepackage{amsmath}
\usepackage{graphicx}
\usepackage{dcolumn}
\usepackage{hyperref}
\usepackage{subfigure}
\usepackage{epsfig}

\newcommand{\beq}{\begin{equation}}
\newcommand{\eeq}{\end{equation}}
\newcommand{\beqa}{\begin{eqnarray}}
\newcommand{\eeqa}{\end{eqnarray}}

\begin{document}

\title{Proportionate growth in patterns formed in the rotor-router model}
\author{Rahul Dandekar and Deepak Dhar}
\affiliation{Department of Theoretical Physics \\Tata Institute of Fundamental Research, Homi Bhabha Road, Mumbai 400005 India}

\begin{abstract}
We study the growing patterns in the rotor-router model formed by adding $N$ walkers at the center of a $L \times L$  two-dimensional square lattice, starting with a periodic background of arrows, and relaxing to a stable configuration. The  pattern is made of large number of triangular regions, where in each region all arrows point in the same direction. The square circumscribing the region, where all the arrows have been rotated atleast one full circle, may be considered as made up of smaller  squares of different sizes, all of which grow linearly with $N$, for $ 1 \ll N < 2 L$. We use the Brooks-Smith-Stone-Tutte theorem relating tilings of squares by smaller squares to resistor networks, to determine the exact relative sizes of the different elements of the asymptotic pattern. We also determine the scaling limit of the  function describing the variation of  number of visits to a site with its position in the pattern. We also present evidence that deviations of the sizes of different small squares from the linear growth for finite $N$ are bounded and quasiperiodic functions of $N$.
\end{abstract}

\maketitle

\section{Introduction}

As baby animals grow from birth to adulthood, different parts of the body grow roughly proportionately to each other. This is called proportionate growth. While many models of growth have been studied in physics before, e.g. diffusion limited aggregation, invasion percolation, formation of snow-flakes etc., none of these show proportionate growth.  In fact, examples of propotionate growth outside biology are hard to find. One cellular automaton model where this property is seen  is  the patterns formed in the  abelian sandpile model (ASM) \cite{ds13}. Here, the  patterns are formed by depositing particles one by one at the origin of a finite square lattice and letting the configuration relax using the ASM relaxation rules,  until it becomes stable. The growth rates and internal structures of patterns formed starting from various periodic backgrounds have been characterized (\cite{dsc09},\cite{sd10},\cite{sd12}). The effect of various forms of noise on the patterns has also been studied (\cite{sd11}).\\

In this paper, we extend these results for pattern formation showing proportionate growth  to a related model - the rotor-router model. This model  shares the abelian property with the sandpile model, and in fact, the operators corresponding to particle addition satisfy the same algebra as the sandpile model \cite{dharsoc}. We characterize the structure of these patterns. The characterization here turns out to be simpler than in the case of the sandpile model, related to the fact that here one deals only with piece-wise linear functions, while in the sandpile problem these functions are in general piece-wise quadratic \cite{ds13}. We show how the relative sizes of different elements in the patterns so formed can be calculated exactly from the knowledge of the adjacency structure of the pattern. We also present numerical evidence that as  $N$ tends to infinity, the deviations of the diameter of the pattern about the linear growth law in $N$ remain bounded, and show interesting quasiperiodic properties.\\

The rotor-router model is a simple model of a deterministic walk, in which the walker locally modifies the medium it moves in, affecting its subsequent motion when it returns to the same site. The model was originally introduced in the context of self-organized criticality, and called the Eulerian Walker model (\cite{pddk},\cite{pps98}).  The name comes from the fact that on a finite undirected graph, the walk eventually settles into an  Euler cycle, in which each edge of the graph is visited exactly once in each direction.  It was independently proposed by Jim Propp as a derandomized version of the random walk, and called  the rotor-router model (\cite{propprev}).  The latter nomenclature seems to describe the model better, and will be used in this paper. Arising from the general interest in derandomization, difference between the fluctuation properties of the rotor-router walk and the the random walk have been the subject of much interest (\cite{holpropp}). Reviews of  earlier work on the rotor-router model may be found in  \cite{mathreview} and \cite{levinethesis}. \\

\begin{figure}[h]
\centering
\subfigure[]{
\includegraphics[width=0.4\linewidth,height=0.38\linewidth]{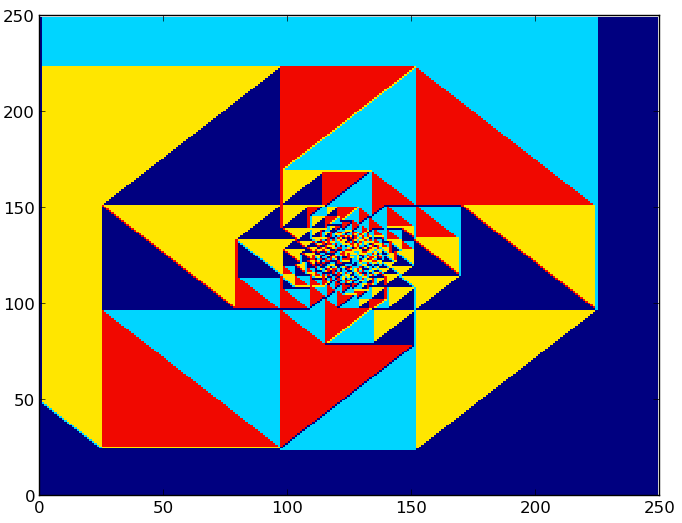}
}
\subfigure[]{
\includegraphics[width=0.4\linewidth,height=0.38\linewidth]{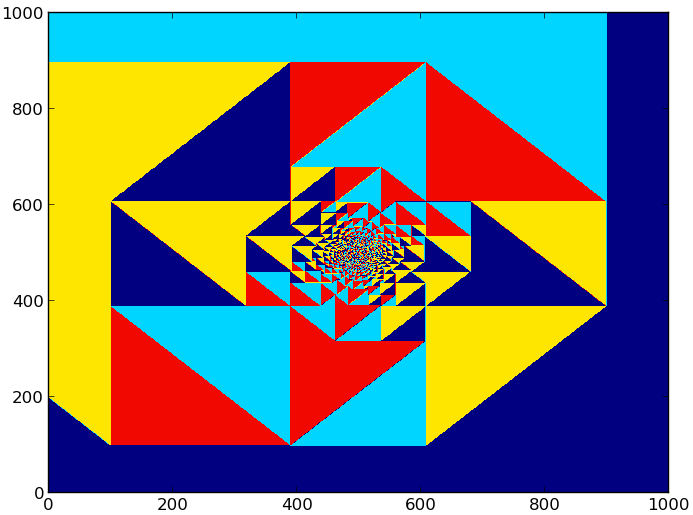}
}
\caption{Pattern formed by depositing rotor-router walkers at the origin of a square lattice, starting from an initial configuration of all arrows pointing East. (a)  the pattern after 400 walkers and (b) after 1600 walkers. Note that the second picture is scaled by a factor 4 compared to the first. Colour code: dark blue - $\rightarrow$, light blue - $\uparrow$, yellow - $\leftarrow$, red - $\downarrow$}
\label{patt}
\end{figure}

If we allow walkers to leave the lattice at the boundaries, and new walkers are added at randomly selected sites, the rotor-router model reaches a critical steady-state which shows self-organized criticality.  Also, like the abelian sandpile model,  if walkers are added at one site, the pattern of arrows obtained after relaxation shows proportionate growth.   The diameter of these patterns and the sizes of their internal elements grow linearly with the number of walkers added.  The simplest pattern of this type  (fig. \ref{patt}) was in fact first observed by Holroyd and Propp in \cite{holpropp}, but these authors were mainly interested in the relation of this problem to random walks. Escape rates of walkers and the behaviour of its diameter as a function of the number of walkers were studied by Florescu et al \cite{escaperates}. They showed that, in $d \ge 2$ dimensions, the diameter of the patterns by depositing $N$ particles on a periodic transient background grows with $N$ as $\beta N$ where $\beta$ is a finite constant depending on the background. The characterization of the patterns itself has not been undertaken so far. This we do in this paper.\\

We briefly mention other related work.  A version of the rotor-router model where the walker stops walking when it visits a previously unvisited site has been studied by Levine and Peres (\cite{lp05}, \cite{lp09}). They found that the aggregate formed by these walkers around the origin on a 2D lattice has a circular shape, similar to that in Internal Diffusion Limited Aggregation, and derived sharp bounds on the fluctuations in the radius of this "Propp circle".  The Propp circle also shows internal structure similar to the patterns we study here, but this structure has not been studied in detail. Some nice pictures of the Propp circle, and related patterns in the sandpiles may be found in \cite{kleber}. Kapri and Dhar studied the pattern produced by a single walker walking on a random background in 2D \cite{kapridhar}, and found numerical evidence  that the asymptotic shape of the pattern is a circle.  They also argued that the average number of  visits to a site $\vec{r}$ upto time $t$ tends to a  scaling function of the form $t^{1/3} g(\vec{r} t^{-1/3})$. \\

The outline of this paper is as follows: We  define the  model  precisely in Section II, and recapitulate some known results.  In section III,  we study the pattern obtained from the initial configuration with all arrows parallel. Using the Brooks-Stone-Smith-Tutte theorem, we construct the resistor network corresponding to the tiling seen in the pattern and determine the sizes of the tiles. We obtain an exact characterization of the asymptotic pattern, and determine the relative sizes of different elements. In section IV, we determine the visit function for the pattern in the large $N$ limit. In section V, we extend the analysis to some other patterns obtained starting from different periodic initial configurations of arrows.  In section VI give evidence that the fluctuations in the diameter as a function of $N$ are bounded and show quasiperiodicity. Section VII contains a summary of our results, and some concluding remarks. In an appendix we rederive the results of section IV without using the tiling picture.

\section{Definition of the Model}\label{def}

\begin{figure}[t]
\begin{center}
\includegraphics[scale=0.5]{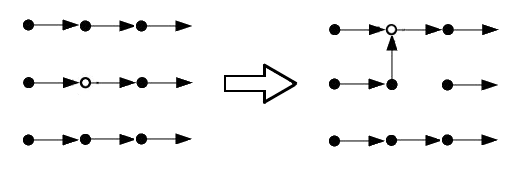}
\end{center}
\caption{A single rotor-router move: The position of the walker is marked by an unfilled circle.}
\label{move}
\end{figure}

We will consider the  rotor-router model on a two-dimensional square lattice. There is an arrow attached to every lattice site, which points in the direction of one of its four neighbours. When a walker reaches a site, it rotates the arrow attached to that site by $90^{\circ}$ counterclockwise and takes a step  in the new direction of the arrow (fig. \ref{move}). For sites which have been visited by the walker, the current direction of the arrow shows the direction of last exit of the walker from that site. Since the update procedure is deterministic, the configuration of arrows and the position of the walker after $n$ steps is fully determined by the initial configuration.\\

We consider a finite square lattice, with the first walker starting at the origin. The arrows attached to the boundary sites can point away from  the lattice. The walker then `falls off' the lattice when it follows such an arrow. When this happens, a new walker is introduced at the origin.\\

We study the configuration of arrows when the $N$-th  walker has just left the system, and the $N+1$-th walker is not yet introduced. First consider the case where the initial configuration is such that all arrows are parallel, pointing due East. Clearly, the first walker put at the origin would move along the positive y-axis,  rotating the arrows it encounters to point North. The second walker rotates the arrow at the origin to point West, and then walks North along the vertical line $x=-1$. The third walker returns to the origin once, before leaving along the vertical line to the left of the previous walkers' paths. And so on.\\

\begin{figure}[h]
\centering
\subfigure[]{
\includegraphics[width=0.3\linewidth,height=0.28\linewidth]{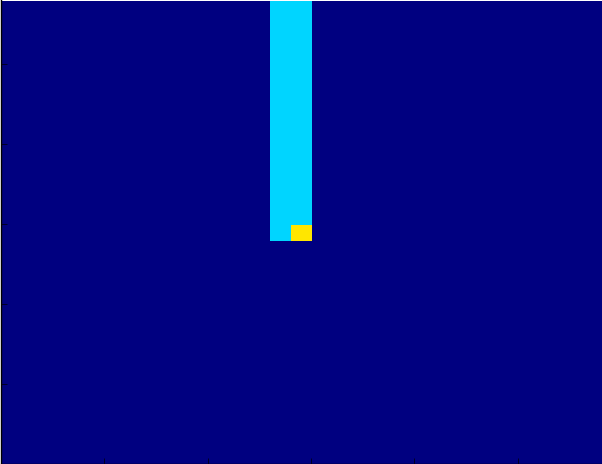}
}
\subfigure[]{
\includegraphics[width=0.3\linewidth,height=0.28\linewidth]{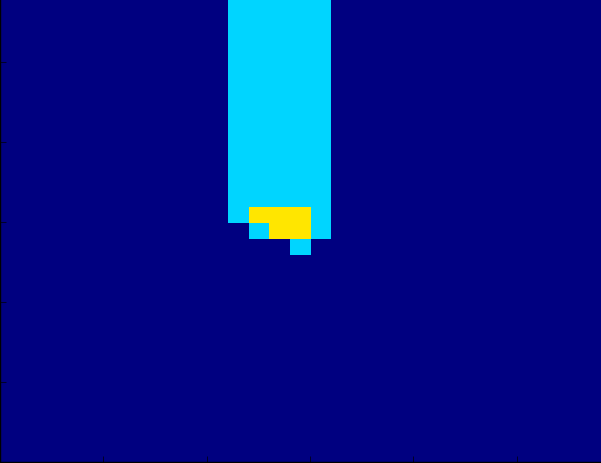}
}
\subfigure[]{
\includegraphics[width=0.3\linewidth,height=0.28\linewidth]{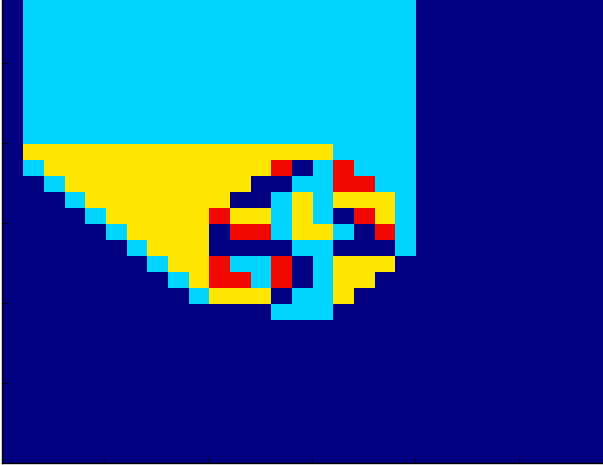}
}
\caption{The arrangement of the arrows starting from the initial background with all spins point East after (a) two (b) five and (c) twenty walkers.  The lattice depicted is $50 \times 50$. colour code: dark blue - $\rightarrow$, light blue - $\uparrow$, yellow - $\leftarrow$, red - $\downarrow$.}
\label{unievol}
\end{figure}

Fig \ref{unievol} shows the configurations left behind by the walker after 2, 5 and 20 walkers. We denote sites having different directions of arrows by  small filled squares of different colours. Then the region of the lattice visited  at least once by a walker is seen to consist of three distinct regions: a dart-shaped region with all arrows pointing west, a vertical stripe in which all arrows are pointed towards north, showing the walkers' paths before they leave from the top  boundary, and a growing octagonal region around the origin. The octagonal region is made of sites where the arrows have undergone atleast one complete rotation, while sites on the rest of the lattice have been visited less than four times. We call this octagonal region as the `pattern'. We first note that the internal structure of the pattern grows proportionately with the pattern itself. This is seen by comparing Fig. \ref{patt} (a) and (b), which shows the pattern after  the number of walkers $ N=400$  and $1600$ respectively. The diameter $D(N)$ of the pattern is defined as the extent of the pattern along the x-axis, after $N$ walkers have exited from the lattice.  We  find that, asymptotically, $D(N) \sim N/2$. The behaviour of $D$ as a function of $N$ is discussed in detail in section \ref{sec:quasi}.\\

It is seen that the octagon  is made of triangular or dart-shaped quadrilateral  regions in which all arrows point in the same direction (shown with the same colour). We call these three- or four-sided polygonal regions `patches of constant orientation'. As $N$ is increased, the size of these patches increases, but their shapes and relative sizes remain unchanged except for inevitable small fluctuations, which however are of $O(1)$ and which we shall study in section \ref{sec:quasi}.\\

\begin{figure}[h!]
\centering
\subfigure[]{
\includegraphics[scale=0.33]{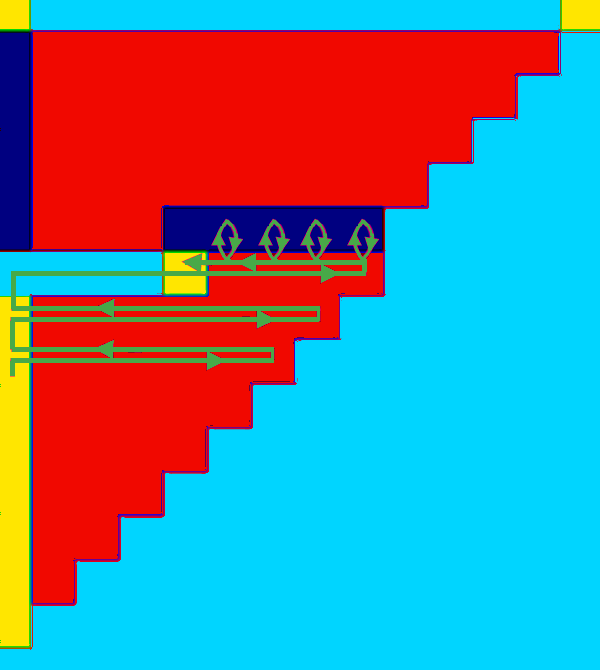}
}
\subfigure[]{
\includegraphics[scale=0.4]{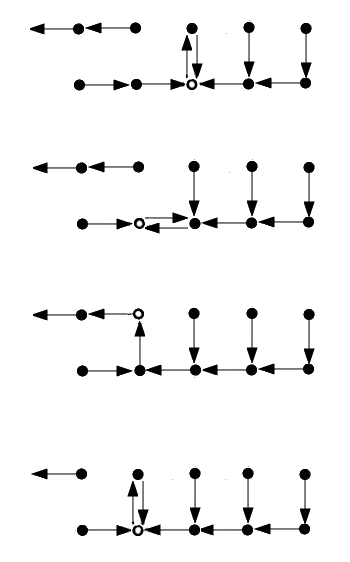}
}
\caption{(a) A schematic representation of the trail of the walker in a single red patch : When going to the right the path is a straight line, while going to the left, it moves between two adjacent horizontal rows, as indicated  in the topmost part of the path.  (b)  Succesive arrow configurations encountered during the walker overall leftward motion. The position of the walker is indicated by an unfilled circle. The colour code: dark blue - $\rightarrow$, light blue - $\uparrow$, yellow - $\leftarrow$, red - $\downarrow$}
\label{onepatch}
\end{figure}

We now describe how a walker explores a single patch after entering it (fig. \ref{onepatch}). From the evolution rules of the walker, it is easy to check that when a walker enters a patch, it visits all the sites in the patch four times, except for sites on the boundaries, before leaving the patch. Thus, the orientation of arrows in the patch does not change with $N$, except at the boundaries of the patch, which may be shifted by one lattice spacing. A new walker, introduced at the origin, moves from one patch to another until it exits the octagonal region, after which it follows a non-self-intersecting path to the sink. As a result of this motion, the boundaries of patches visited shift, and the overall pattern  grows in size.\\

\section{Characterizing the Pattern as a Tiling}

In this section we will determine the sizes of the different patches in the pattern, using as input the observed  arrangement of these elements in the pattern. We do this by  noting that the observed pattern may be thought of as a realization of tiling of a square by smaller squares, where each square tile is made of two triangles of different colours (compare figs. \ref{patt} and \ref{patt_til} (a)) . We will ignore the colours, and focus on the sizes and the adjacency relations between different tiles.\\

 \subsection{The Brooks-Smith-Stone-Tutte Mapping}

\begin{figure}[t]
\begin{center}
\includegraphics[scale=0.45]{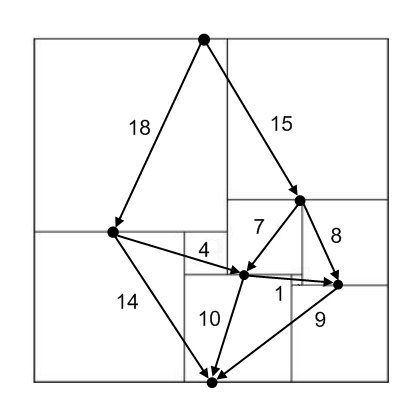}
\end{center}
\caption{A tiling of a 33x32 rectangle by squares and the corresponding resistor network. All the edges have resistance unity and the arrows show the direction of the current.}
\label{bsst}
\end{figure}

The Brooks-Smith-Stone-Tutte (BSST) mapping (\cite{bsst}) between square tilings and resistor networks allows one to find the sizes of all the squares in a square tiling using their arrangement in the tiling. We first recapitulate the BSST mapping and then use it to calculate the properties of the pattern in fig. \ref{patt}.\\

Consider the tiling of a rectangle by squares shown in  Fig. \ref{bsst}. We associate with a given tiling a resistor network as follows: Consider horizontal segments in the tiling as nodes of the circuit, and join two nodes by a ($1 ~ \Omega$) resistor if parts of the corresponding segments form opposite edges of  a tiling square. The resistors (that is, the edges) in the network thus correspond to squares. We associate a voltage $V(v)$ with the nodes $v$ of the network, where $V(v)$ is the vertical height of the horizontal edge corresponding to the node $v$ in the tiling. Assigning unit resistance to all resistors, the current between nodes $v$ and $u$ is $V(v) - V(u)$, and this equals the length of the side of the square joining the corresponding horizontal lines in the tiling. Then the Kirchoff current balance conditions of the resistor network corresponds to the condition that the total horizontal length $L(v)$ of the segment $v$ is the same whether calculated using its set of upper neighbours $\{u\}$ or the set of lower neighbours $\{ \ell \}$

\beqa
\sum_{\{u\}} [V(u) -V(v)] = \sum_{\{ \ell \}} [V(v) -V(\ell)] = L(v) \mbox{~~~ or,}\\
\sum_{n} [V(n) -V(m)] = 0
\eeqa

where the summation in the second line is over all neighbors $n$ of the node $m$ in the graph. Also, $V(U)=0$ and $V(L)=1$ where $L$ and $U$ are the vertices corresponding to lower and upper edges of the bounding square. Then, the unique solution for the voltage developed on the nodes of the resistor network constructed in the above way gives the heights (in the original tiling) of all the horizontal segments.\\

Note that when four squares meet at a point, the horizontal segment going through the point can be considered as one segment or two.  An  example is the meeting point of  segments G and B in fig. \ref{patt_til} (a)). In such cases, we will always choose the latter, more general, option. (The former case corresponds to the degenerate case when the voltages at B and G are equal.) 

\subsection{The Pattern as a Square Tiling}

\begin{figure}[h!]
\centering
\subfigure[]{
\includegraphics[scale=0.45]{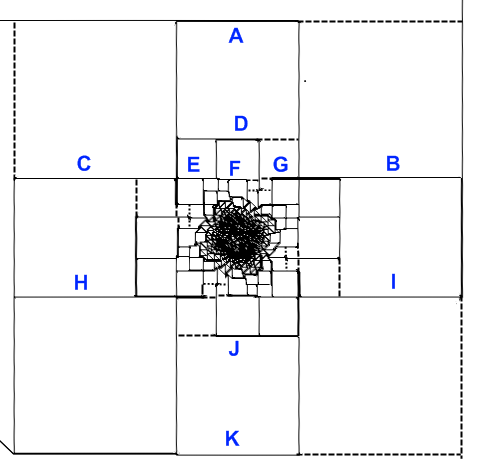}
}
\subfigure[]{
\includegraphics[scale=0.45]{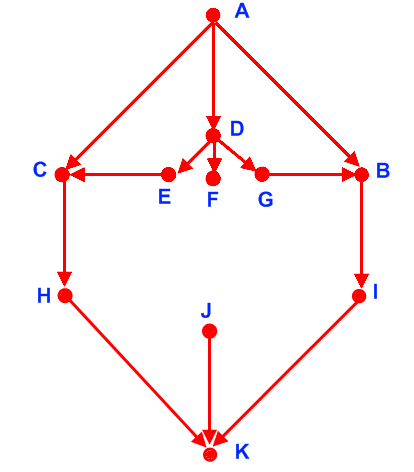}
}
\subfigure[]{
\includegraphics[scale=0.4]{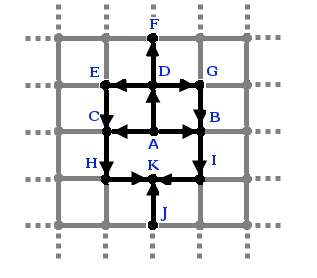}
}
\caption{(a) The pattern as a square tiling. The first few levels of the corresponding network are shown. (b) Part of the resistor network corresponding to the tiling above. The edges illustrated part (a) are in black with the direction of the current shown as it was previously.  Note that some nodes (for example (1,0) and (1,1)) correspond to horizontal segments which lie at the same height, but have not been grouped together as a single segment. (c) The graph in part (b) shown as a part of the complete resistor network for the tiling in (a), which is a square lattice.}
\label{patt_til}
\end{figure}

We consider the pattern created starting from a uniform starting background (Fig. \ref{patt}). Since the diameter of this pattern grows as $D = N/2$, we rescale the pattern by $N/2$ and take $N \rightarrow \infty$ to define an rescaled `asymptotic' pattern with diameter $D' = 1$.
This pattern has  features at arbitrarily small scales. Fig. \ref{patt_til} (a)  shows the pattern as a tiling of the unit square by smaller squares. The corresponding resistor network to three layers starting from the outer boundary is shown in \ref{patt_til} (b). Drawing more layers of the resistor network, one will get an infinite graph with an infinite number of nodes. Most of these nodes correspond to the very tiny square tiles near the center of the pattern.\\

The main simplification that allows us to analyse this problem exactly is the fact that the resistor network graph has a very simple structure :  it is equivalent to the resistor-network formed by connecting unit resistors to form a square grid, with only one missing bond between nodes A and K (Fig. \ref{patt_til} (c)). One can easily verify that  the networks in (b) and (c) are equivalent, being different pictorial representations of the same graph. For the present, we will take this crucial observation as an induction from the features of the observed pattern. An explanation of why the resistor network is a square grid follows from the consideration of the visit function in Sec. IV.\\

For the resistor network shown in Fig. \ref{patt_til} (c),  it is convenient to label nodes by their integer coordinates $(m,n)$. We will choose the coordinates so that the node  A  has the Euclidean coordinate $(0,0)$ and the node K  has the coordinate  $(0,-1)$.  The boundary conditions for the voltage on this resistor network are given by the positions of the upper and lower horizontal segments of the bounding square. Since the pattern has been rescaled to be of unit diameter centred at the origin, this gives $V(0,0) = 1/2$, and $V(0,-1)=-1/2$.\\

The solution of the resistor network problem for this graph is well known, $V(m,n)$ being given by the following superposition of Green's functions for an infinite square lattice $G_{sq}(m,n)$

\beq
V(m,n) = 2 (G_{sq}(m,n+1) - G_{sq}(m,n))
\label{eq:Vmn}
\eeq
where
\beq
G_{sq}(m,n) =    \int_{-\pi}^{\pi} \frac{dk_1}{2 \pi} \int_{-\pi}^{\pi} \frac{dk_1}{2 \pi}  \frac{ 1 - \cos ( k_1 m + k_2 n)}{2 - \cos k_1 - \cos k_2}
\eeq

It can be verified that the normalisation is indeed $V(0,0) = 1/2$. For finite values of $m$ and $n$, this integral can be evaluated in closed form, and the result is of the form $a_{m,n} + b_{m,n}/ \pi$, where $a_{m,n}$ and $b_{m,n}$ are rational numbers \cite{spitzer}. A computationally efficient formula for $G_{sq}(m,n)$, which can be used in Mathematica$^{TM}$ to get exact expressions for the first few $V(m,n)$, is given in \cite{cserti},

\beq
G_{sq}(m,n) = \int_{0}^{\pi} \frac{dy}{2 \pi} \frac{1 - e^{|m|s} \cos (ns)}{\sinh (s)} \label{eq:Gsq}
\eeq

with $\cosh (s) = 2 - \cos (y)$. \\

From this solution we get the sizes of various elements in the pattern. For example, the size of the big squares at the four corners of Fig. \ref{patt_til} (a) is given by the difference in the vertical co-ordinates of lines A and B. From Fig. \ref{patt_til} (c) this is equal to $V(A)-V(B) = V(0,0) - V(1,0)$. Using the values $V(1,0) = \frac{2}{\pi} - 1/2$ and $V(0,0) = 1/2$, the size of these largest squares in the pattern relative to the size of the pattern is $1-\frac{2}{\pi}$.\\

For $m,n \gg 1$, the corresponding horizontal lines get closer and closer to the origin of the original pattern. On the resistor network for large $m,n$, $V(m,n)$ looks like the electric potential due to a dipole at the origin, and tends to zero for large $m,n$. $V(m,n) \sim \frac{\cos{\theta}}{(m^2+n^2)^{1/2}} = \frac{n}{(m^2+n^2)^{3/2}}$ in this region. Consider moving alone the y-axis of the resistor network. Then $V(m=0,n) \sim 1/n$. In the tiling, this gives the distance from the origin of the n$^{th}$ ring of squares (counted from the outside inwards) is $r_n \sim 1/n$. Then the size of the squares in this ring varies as $\frac{\partial r_n}{\partial n} \sim \frac{1}{n^2}$.

\section{The Visit Function} \label{sec:visit}

For the sandpile patterns, the exact characterization of the patterns was given in terms of the toppling function  $\phi_N(x,y)$ which gives the number of topplings at any lattice point $(x,y)$, when $N$ particles are added at the origin, and the configuration is relaxed. For the rotor-router model, the corresponding function, which we also denote $\phi_N(x,y)$ is the so-called visit function, which counts the total number of full rotations undergone by the arrow attached to the site $(x,y)$ due to the first $N$ walkers. The region where $\phi_N$ is non-zero is just the octagonal region which we have earlier defined as our pattern. We define scaled coordinates $\xi =x/D_N, \eta = y/D_N$, where $D_N$ is the diameter of the pattern after $N$ walkers have exit the lattice. Define the scaled visit function $\phi(\xi,\eta)$ by

\beq
\phi(\xi, \eta) = \lim_{N \to \infty} \frac{\phi_N(\lfloor D_N\xi \rfloor,\lfloor D_N\eta \rfloor)} {D_N} \label{eq:scaled}
\eeq

The scaling implies that upper and lower boundaries of the scaled pattern lie along $y=1/2$ and $y=-1/2$, as in the previous section. We now characterize the behaviour of $\phi$.\\

\begin{figure}[t]
\centering
\subfigure[]{
\includegraphics[scale=0.5]{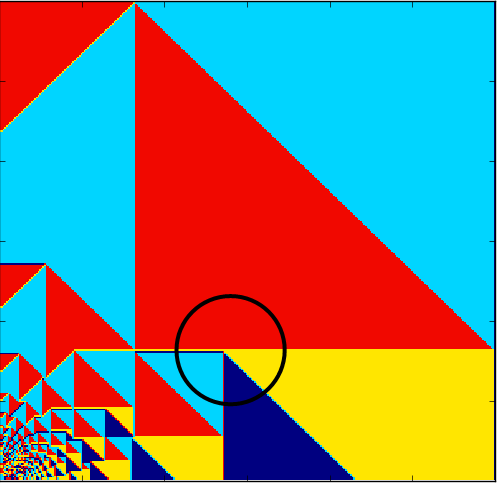}
}
\subfigure[]{
\includegraphics[scale=0.5]{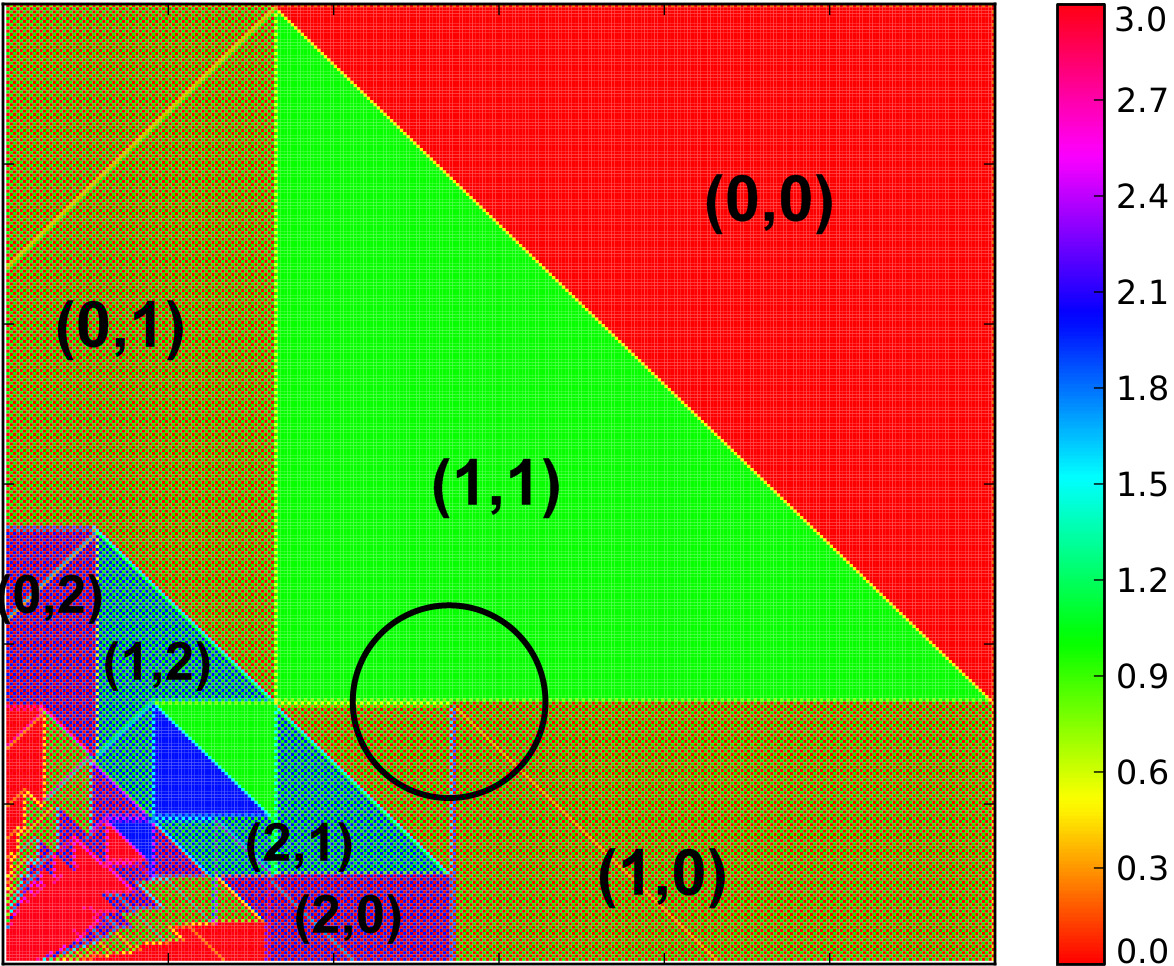}
}
\caption{(a) The first quadrant of the pattern in Fig. 1, and (b) the the same area showing the patches in the visit function. What is being plotted in (b) is the function $\nabla \phi_N$ on the lattice, with the x-component plotted on the odd sites of the lattice and the y-component on the even sites. The values of the vector $\nabla \phi_N$ for some of the patches are shown. Note that there are four patches of different arrow orientation within the circle shown in (a), whereas only two regions with different values of $\nabla \phi_N$ in the corresponding circle in (b). The color code in (a) is: dark blue - $\rightarrow$, light blue - $\uparrow$, yellow - $\leftarrow$, red - $\downarrow$.}
\label{patchespatches}
\end{figure}

\begin{figure}[t]
\centering
\subfigure[]{
\includegraphics[scale=0.42]{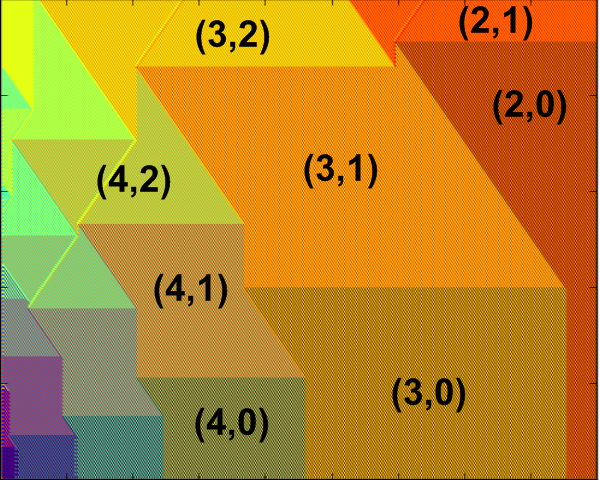}
}
\subfigure[]{
\includegraphics[scale=0.25]{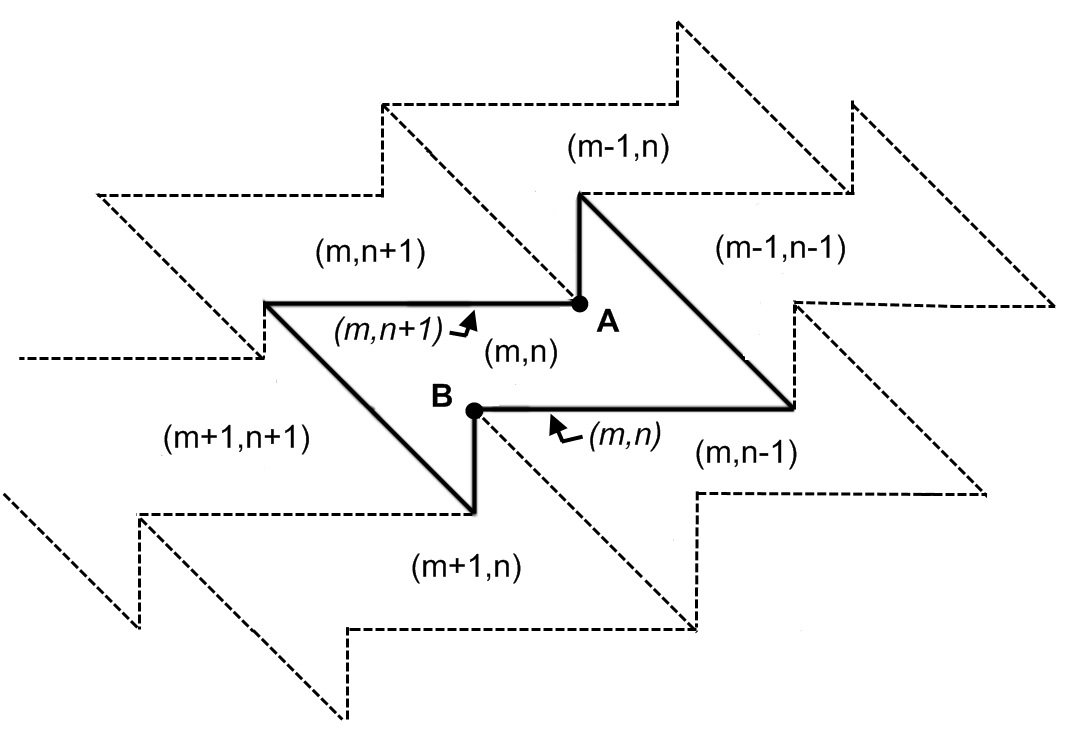}
}
\caption{ (a) A zoomed-in figure of some patches in Fig. 9 (b), illustrating the arrangement of patches of the visit function, labeled by their $(m, n)$ values. Note that the vector $\nabla \phi_N$ changes by $(0,1)$ across a horizontal boundary, by $(1,0)$ across a vertical one, and by $(1,1)$ across a diagonal boundary. (b) The patch $(m,n)$ and its neighbours. The points A and B lie on a line of slope $1/2$. The lower horizontal boundary of the patch is given the same co-ordinate as that of the patch.}
\label{visitpatches}
\end{figure}

$\phi(\xi, \eta)$ is a continuous function of both variables. Using induction on the way the visit function changes with $N$ (refer Fig. \ref{onepatch}), it is easily seen that within a given patch, $\nabla \phi_N$ is a constant independent of $N$. However, its value might change discontinuously across patch boundaries (but it does not always, as is shown in fig. \ref{patchespatches}). When it is non-zero, this discontinuity in $\nabla \phi_N$ across a particular boundary might be $(0,1)$, $(1,0)$ or $(1,1)$, across horizontal, vertical and diagonal boundaries respectively. However, sometimes two adjacent patches with different arrow orientations might have the same value of $\nabla \phi_N$. Hence we define `patches of the visit function' as the regions across which $\nabla \phi_N$ is the same constant. To compare patches of the visit function with patches of constant orientation, refer Fig. \ref{patchespatches} (a) and (b).\\

We now {\emph prove} that $\nabla \phi$ has to be a constant within patches for a linearly growing pattern \cite{ds13}. That is, we prove that within each patch, $\phi(x,y)$ behaves as $f_{per} (x,y) + f D_N + v_1 x + v_2 y$ where $f_{per}$ is a periodic function on the lattice and $f$, $v_1$ and $v_2$ are constants which depend on the patch.\\

\begin{figure}[t]
\begin{center}
\includegraphics[scale=0.4]{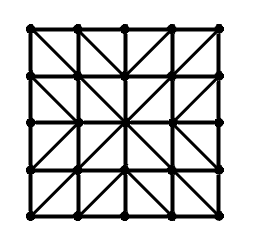}
\end{center}
\caption{A part of the  adjacency graph of the visit function, corresponding to the patches shown in Fig. 3(b). The central node is the vertex $(0,0)$ corresponding to the outer region not visited by the walkers }
\label{adj}
\end{figure}

To prove this, consider the scaled toppling function $\phi(\xi, \eta)$ within a patch of a single colour. Expanding $\phi$ about a point, one gets to get $\phi = f_0 + v_1\eta + v_2\xi + v_3\eta^2 + v_4\xi^2 + \dots$ This implies that the rescaled visit function $\phi_N$ behaves near the point $(x,y) = (\lfloor D_N\xi \rfloor,\lfloor D_N\eta \rfloor)$ as $\phi_N = f + f_0 D_N + v_1x + v_2y + (v_3x^2 + v_4y^2)/D_N + \dots$, where $f$ is a constant of order unity. Since $\phi_N$ is an integer valued function, $v_3, v_4$ and all higher coefficients have to be zero. Therefore, within each patch, $\phi(\xi,\eta)$ is of the form stated earlier

\beq
\phi(\xi,\eta) = f + v_1 \xi + v_2 \eta
\eeq

Within a single patch, $\delta \phi_N(x) = \phi_N(x+\delta x) - \phi_N(x) = (v_1,v_2) \cdot \delta x$ has to be an integer, where $\delta x$ is the basis vector of the superlattice of the pattern within the patch. For the present pattern, the periodicity in both directions is $1$. Hence $v_1,v_2$ can only take integral values. Thus, within a patch, $\phi_N(\xi,\eta)$ is of the form

\beqa
\phi_N(x,y) &=& f_{m,n} + n x + n y
\eeqa

where $m$ and $n$ are integers depending on the patch. For the present pattern, as discussed previously, $m,n$ only change by $\pm 1$ across horizontal or vertical boundaries. This implies that patches with all allowed $m,n$ values are present. The adjacency graph for patches of the visit function is shown in Fig. \ref{adj}. Figs \ref{visitpatches} (a) shows the arrangement of these patches in the actual figure.\\

To determine the complete visit function we have to determine $f_{m,n}$ for each patch. To do this we use the results of the previous section for the tiling.\\

First we choose a rule to assign a horizontal segment to each visit function patch. Each patch has two horizontal boundaries. To each patch, we associate  the horizontal segment of the tiling which goes through its lower boundary. This association is one-to-one, and chosen such that both the patch and the segment associated with it can be labelled with the same indices $(m,n)$. Fig \ref{visitpatches} (b) shows the arrangement of a patch of the visit function along with it's neighbouring patches, and the corresponding horizontal lines. For two patches with co-ordinates $(m,n)$ and $(m,n+1)$, the corresponding visit functions are given by $\phi_1(\xi,\eta) = m\xi + n\eta + f_{m,n}$ and $\phi_2(\xi,\eta) = m\xi + (n+1)\eta + f_{m,n+1}$. The continuity of the visit function along the horizontal boundary between patches $(m,n)$ and $(m,n+1)$ gives the position of the boundary as

\beqa
y = \eta_{m,n} \mbox{~~~~ with}\\
\eta_{m,n} = f_{m,n} - f_{m,n+1}
\eeqa

Comparing with equation \ref{eq:Vmn}, we can make the identification 

\beq
f_{m,n}= -2 G_{sq}(m,n) \label{eq:fmn}
\eeq

This gives $f_{0,0} = 0$ and $f_{0,1} = -1/2$, which is consistent with the normalisation that the upper boundary of the pattern (the boundary between patches $(0,1)$ and $(0,0)$ lies along the line $x=1/2$. The $f$'s are negative because patches with positive $m,n$ lie in the upper right quadrant of the pattern. For large $m$ and $n$, $f_{m,n} \sim -\log{(m^2+n^2)}$.\\

Appendix A gives a derivation of eqn. (\ref{eq:fmn}) without using the tiling picture.

\section{Other starting backgrounds}\label{sec:other}

\begin{figure}[h]
\centering
\subfigure[]{
\includegraphics[scale=0.25]{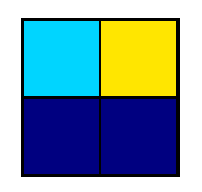}
}
\subfigure[]{
\includegraphics[width=0.4\linewidth,height=0.37\linewidth]{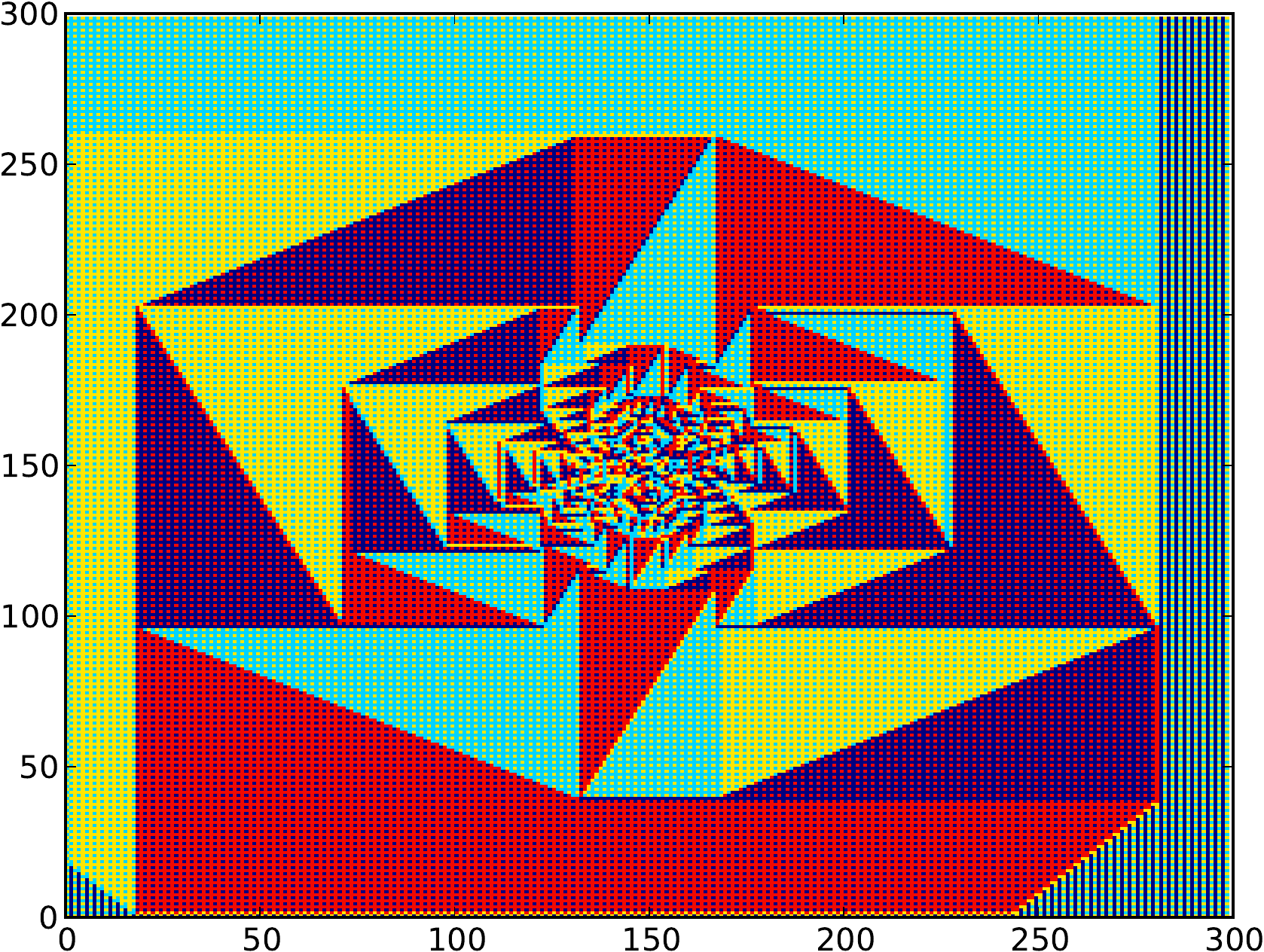}
}
\subfigure[]{
\includegraphics[scale=0.33]{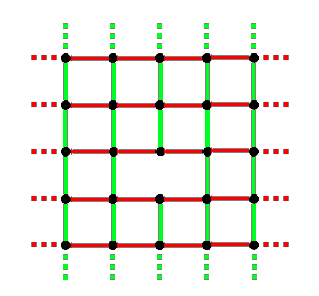}
}
\caption{Pattern formed the initial backgrounds generated by 2x2 unit cell given in (a) is shown in (b) after 700 walkers put at the origin have left the lattice. Colour code: dark blue - $\rightarrow$, light blue - $\uparrow$, yellow - $\leftarrow$, red - $\downarrow$. (c) The resistor network for the pattern, where black bonds denote a resistance of 1 $\Omega$, red that of $0.5 \Omega$ and green a resistance of $2 \Omega$.}
\label{othera}
\end{figure}

In this section we study the patterns formed by the walker starting with periodic backgrounds formed by periodic repetition of unit cells of size 2. All patterns formed starting from such backgrounds are one of the four types - the following three and the one in fig. \ref{patt}. \\

Let us first consider the pattern formed starting from the unit cell in fig. \ref{othera} (a), shown in fig. \ref{othera} (b) . This pattern can be seen as tilings of the unit square by rectangles with aspect ratio 1:2 . The BSST mapping is easily generalised to a tiling by rectangles: the resistor in the network corresponding to a rectangle with vertical to horizontal side ratio $p/q$ has resistance $R = p/q$. The corresponding resistor networks as drawn in Fig. \ref{othera} (c). Interestingly, this network has the same structure as the one for the pattern studied in section IV, with different resistances along the bonds depending on the aspect ratio of the tile. The vertical bonds have $R_1 = 2 \Omega$ and horizontal bonds have $R_2 = \frac{1}{2} \Omega$. The solution for the potential is given by the superposition of two Green's functions:

\beq
V(m,n) = Q(G_{rect}(R=4;m,n+1)-G_{rect}(R=4;m,n))\label{eq:rectV}
\eeq

Where $G_{rect}(R;m,n)$ is the lattice Green's function on the rectangular lattice with resistance $R~~ \Omega$ along the vertical bonds and $1 \Omega$ along the horizontal bonds. This lattice is equivalent to the present lattice provided all resistances are scaled down by a factor 2; here this is built into the scale factor $Q$ in eqn. (\ref{eq:rectV}). The computationally efficient formula for $G_{rect}(m,n)$ is (\cite{cserti})

\beq
G_{rect}(R;m,n) = \int_{0}^{\pi} \frac{dy}{2 \pi} \frac{1 - e^{|m|s} cos (ns)}{sinh(s)}
\eeq

where $s$ is defined implicitly $cosh(s) = 1 + \frac{1}{R} - \frac{1}{R}cos y$. Using this formula, $V(m,n)$ can be evaluated exactly using eqn. (\ref{eq:rectV}) and Mathematica$^{TM}$.\\

The charge $Q$ is determined by the condition that $V(0,0) = \frac{1}{2}$, which gives $Q = (4 - \frac{4}{\pi}tan^{-1}2)^{-1}$. The height of the big rectangles at the corners \emph{relative to the height of the figure} is hence given by $l = V(0,0)-V(0,1)  = \frac{1}{2} - \frac{Q}{2}(1-\frac{2}{\pi}tan^{-1}\frac{1}{2}) \approx 0.51857$.\\

We can also determine the aspect ratio of the figure, by also using the BSST mapping for the pattern rotated by $90^{\circ}$. The structure of the resistor network is again the same (Fig. \ref{othera} (c)), except that the $0.5 \Omega$ and $2 \Omega$ bonds are interchanged. Hence, for the new network, $V'(m,n) = V(n,m) \frac{Q'}{Q}$, where $Q'$ is determined by the condition that $V(0,0) = \frac{1}{2}$. This gives $Q' = (1-\frac{2}{\pi}tan^{-1}\frac{1}{2})^{-1}$. One can then calculate the breadth of the big rectangle at the corner \emph{relative to the width of the figure} as $b' = \frac{1}{2} - \frac{Q'}{2}(4-\frac{4}{\pi}tan^{-1}2) \approx 0.86866$. One then uses the condition that $2 l H = b' W$ to determine the aspect ratio of the figure to be $H/W \approx 0.83755$.\\

\begin{figure}[h]
\centering
\subfigure[]{
\includegraphics[scale=0.25]{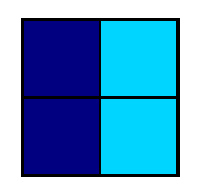}
}
\subfigure[]{
\includegraphics[width=0.4\linewidth,height=0.37\linewidth]{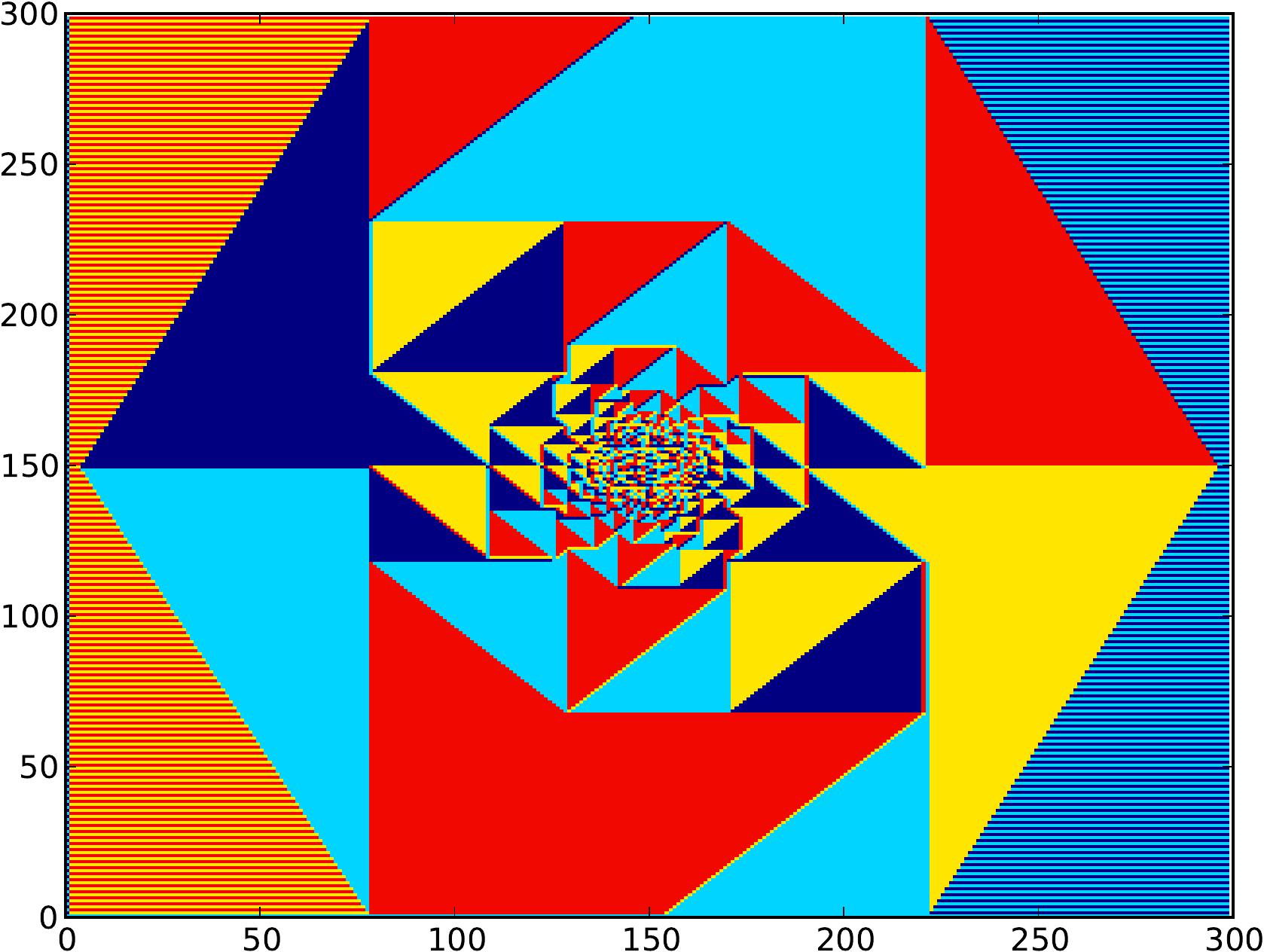}
}
\caption{Pattern formed the initial backgrounds generated by 2x2 unit cell given in (a) is shown in (b) after 900 walkers put at the origin  have left the lattice. Colour code: dark blue - $\rightarrow$, light blue - $\uparrow$, yellow - $\leftarrow$, red - $\downarrow$.}
\label{otherb}
\end{figure}

\begin{figure}[h]
\centering
\subfigure[]{
\includegraphics[scale=0.33]{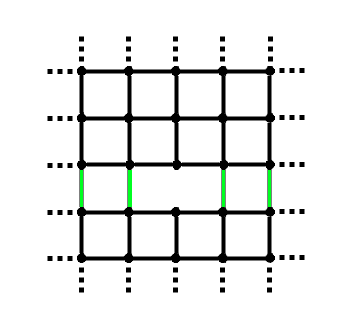}
}
\subfigure[]{
\includegraphics[scale=0.33]{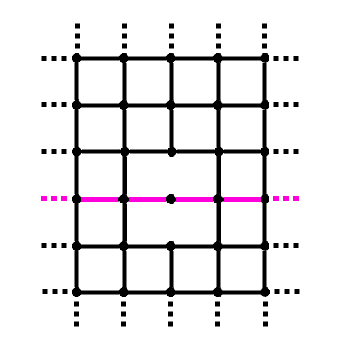}
}
\caption{(a) The resistor network for the pattern in fig. \ref{otherb} (b), where black bonds denote a resistance of 1 $\Omega$, red that of $0.5 \Omega$ and green a resistance of $2 \Omega$. (b) The resistor networks shown in part (a) can be simplified to this, with unit resistances along all bonds. The voltages are constrained to be $\pm \frac{1}{2}$ on the sites shown. The bonds shown in pink are not present in the original networks, being open. However, since in these equivalent networks they connect nodes which have the same voltage, a $1 \Omega$ resistor can be connected between them without affecting the network.}
\label{otherb1}
\end{figure}

Let us now consider the initial periodic structure having the unit cell in fig. \ref{otherb} (a). The resulting pattern after $900$ walkers is shown in fig. \ref{otherb} (b) . The corresponding resistor network is given in fig. \ref{otherb1} (a). It is also a square lattice, but the bonds connecting sites $(i,0)$ and $(i,-1)$ have $R' = 2 \Omega$ whereas the rest of the bonds have $R = 1 \Omega$. Fig. \ref{otherb1} (b) gives an equivalent square lattice structure for this network.\\

Using the BSST mapping, the vertical positions of horizontal lines with respect to the centre of the figure in the upper left half of Fig. \ref{otherb} (b) are given by the voltages in the first quadrant in the square lattice in Fig. \ref{otherb1} (b), taking the point with $+ \frac{1}{2}$ as the origin. In terms of the square lattice Green's function $G_{sq}(m,n)$, eqn. (\ref{eq:Gsq}),

\beq
V_1(m,n) = Q_1 (G_{sq}(m,n+2) - G_{sq}(m,n))
\eeq

$Q_2$ is determined by the constraint $V(0,0) = \frac{1}{2}$ to be $Q_2 = (2 - \frac{4}{\pi})^{-1}$. To determine the aspect ratio of the figure, one determines the positions of the lines with $(m,n) = (0,1)$ and $(1,0)$, call them $l_1$ and $l_2$. The size of the big squares at the boundary is then $(\frac{1}{2}-l_2)$ and that of the square between then is $(\frac{1}{2}-l_1)$. The horizontal extent of the pattern is then $(\frac{1}{2}-l_1) + 2(\frac{1}{2}-l_2) = (\pi - 2)^{-1}$. Since the voltages are normalized such that the vertical extent of the pattern is $1$, This gives the aspect ratio as $H/W = (\pi - 2)^{-1}$.\\

\begin{figure}[h]
\centering
\subfigure[]{
\includegraphics[scale=0.25]{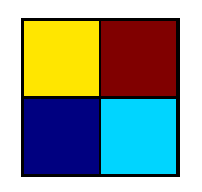}
}
\subfigure[]{
\includegraphics[width=0.35\linewidth,height=0.33\linewidth]{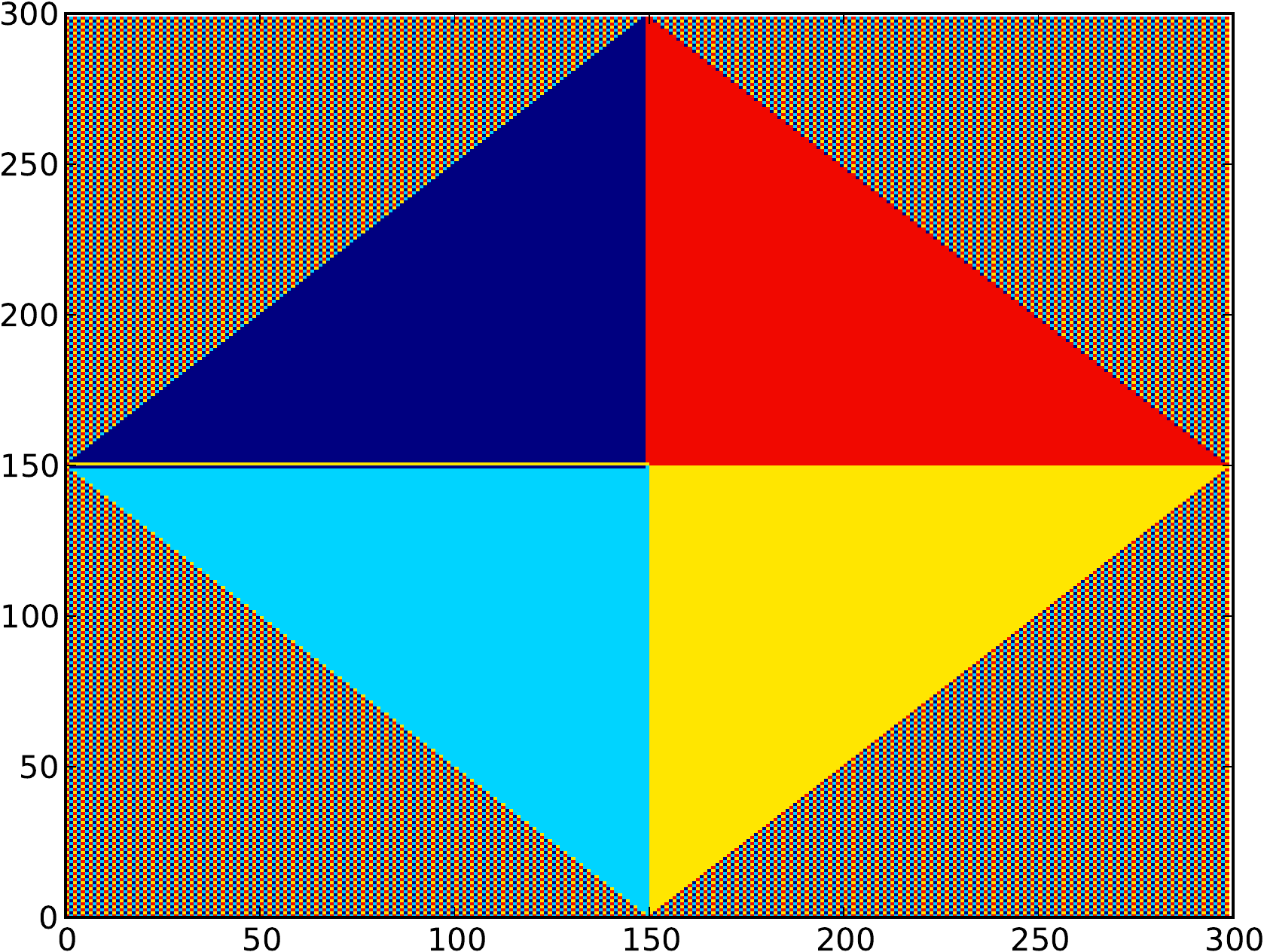}
}
\subfigure[]{
\includegraphics[width=0.35\linewidth,height=0.33\linewidth]{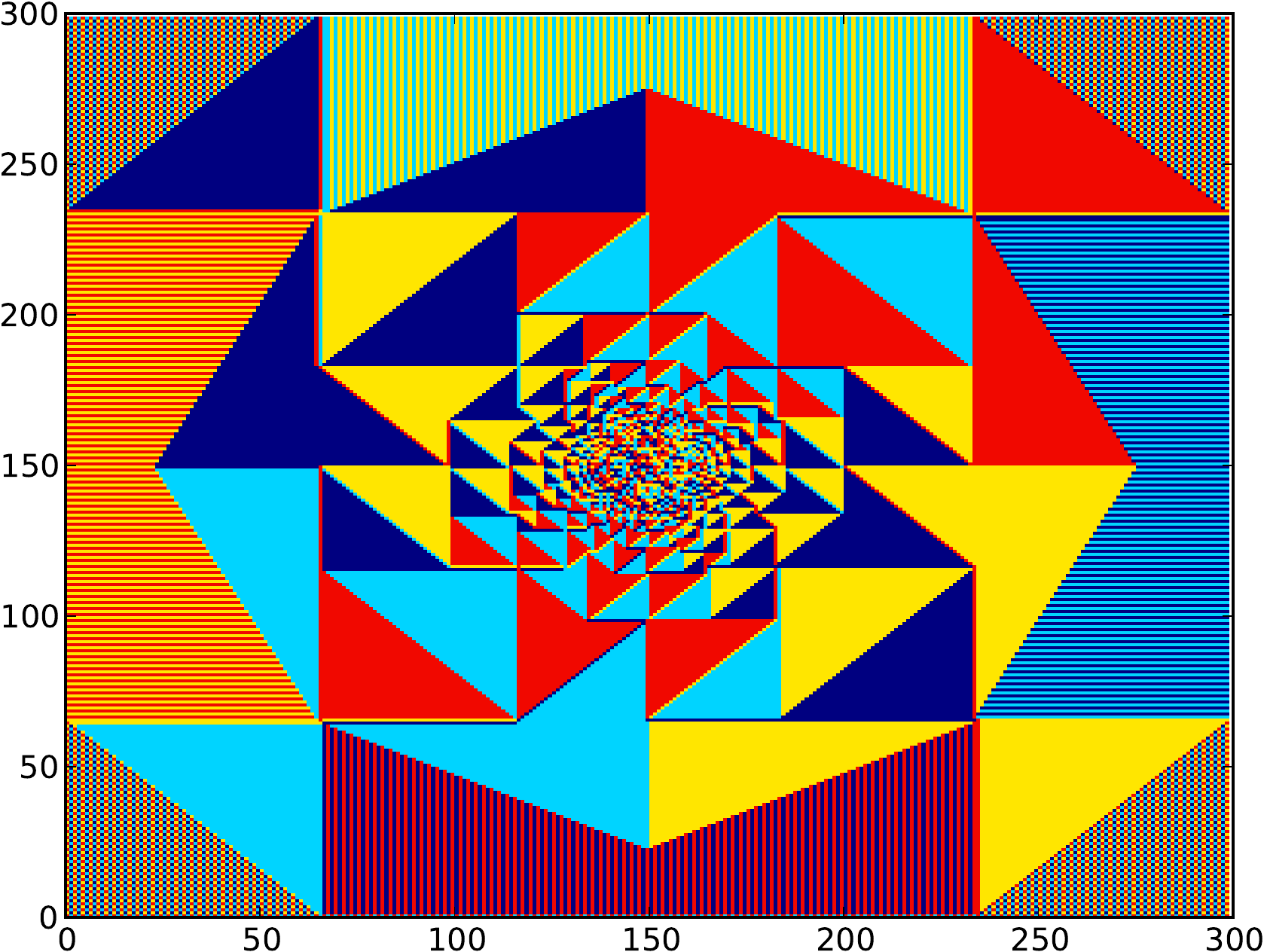}
}
\caption{The pattern formed starting from the (recurrent) initial background shown in (a) by (b) a single walker (c) 1200 walkers on a 600x600 lattice. Colour code: dark blue - $\rightarrow$, light blue - $\uparrow$, yellow - $\leftarrow$, red - $\downarrow$.}
\label{recur}
\end{figure}

\begin{figure}[h]
\centering
\subfigure[]{
\includegraphics[scale=0.33]{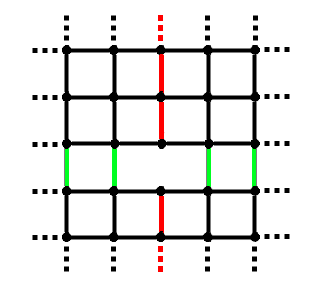}
}
\subfigure[]{
\includegraphics[scale=0.33]{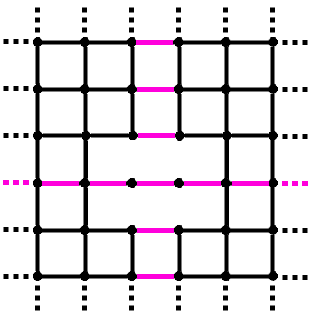}
}
\caption{ (a) The resistor network for the pattern in fig. \ref{recur} (c), where black bonds denote a resistance of 1 $\Omega$, red that of $0.5 \Omega$ and green a resistance of $2 \Omega$. (b) The resistor networks shown in part (a) can be simplified to this, with unit resistances along all bonds. The voltages are constrained to be $\pm \frac{1}{2}$ on the sites shown. The bonds shown in pink are not present in the original networks, being shorted. However, since in these equivalent networks they connect nodes which have the same voltage, a $1 \Omega$ resistor can be connected between them without affecting the network.}
\label{recur1}
\end{figure}

Now we consider the initial periodic structure having the unit cell in fig. \ref{recur} (a). Here the first walker takes of $O(L^3)$ steps to get to the sink, instead of $O(L)$ steps as in the case of the previous patterns. Fig. \ref{recur} (b) shows the pattern left behind by the first walker. Putting more walkers at the origin after the first walker has exit, one gets the pattern observed in Fig. \ref{recur} (c).\\

The resistor network for this tiling is shown in fig. \ref{recur1} (a), and is again a square lattice with bonds with not all resistance equal to $1 \Omega$. Thus, at least starting from 2x2 unit cells, we find that the adjacency graph of patches is the same, and the corresponding resistor networks differ not in the connectivity of the resistors, but only in the values of resistances. Fig. \ref{recur1} (b) gives an equivalent square lattice structure for this network.\\

The voltages in the first quadrant of \ref{recur1} (a) are given by   

\beqa
V_2(m,n) = Q_2 (G_{sq}(m,n-2) &+& G_{sq}(m+1,n-2)\nonumber\\
	 - G_{sq}(m,n) &-& G_{sq}(m+1,n))
\eeqa

where $Q_2$ is determined by the condition that $V(1,0) = \frac{1}{2}$. The aspect ratio for this pattern can be observed from the symmetry of the pattern itself to be unity.

\section{Bounded Fluctuations and Quasiperiodicity} \label{sec:quasi}

In this section we examine the growth rates of the full pattern in fig. \ref{patt} and of the different patches which comprise it, and also the fluctuations about average rates of growth. As a simple example, consider the horizontal extent (distance from the origin) of the pattern along the positive x-axis. Denote this extent after $N$ walkers have left the lattice,  by $H(N)$. Similarly, denote by $H'(N)$ the distance from the origin of vertical line which is the inner border of the patch $(0,1)$ (refer fig. \ref{patchespatches}). In accord with the calculations of section III, we expect that, for large $N$,

\beqa
H(N) &=& \frac{N}{4} + o(N)\\
H'(N) &=& (\frac{3}{4}-\frac{2}{\pi})N + o(N)
\eeqa

\begin{figure}[h]
\centering
\includegraphics[scale=0.4]{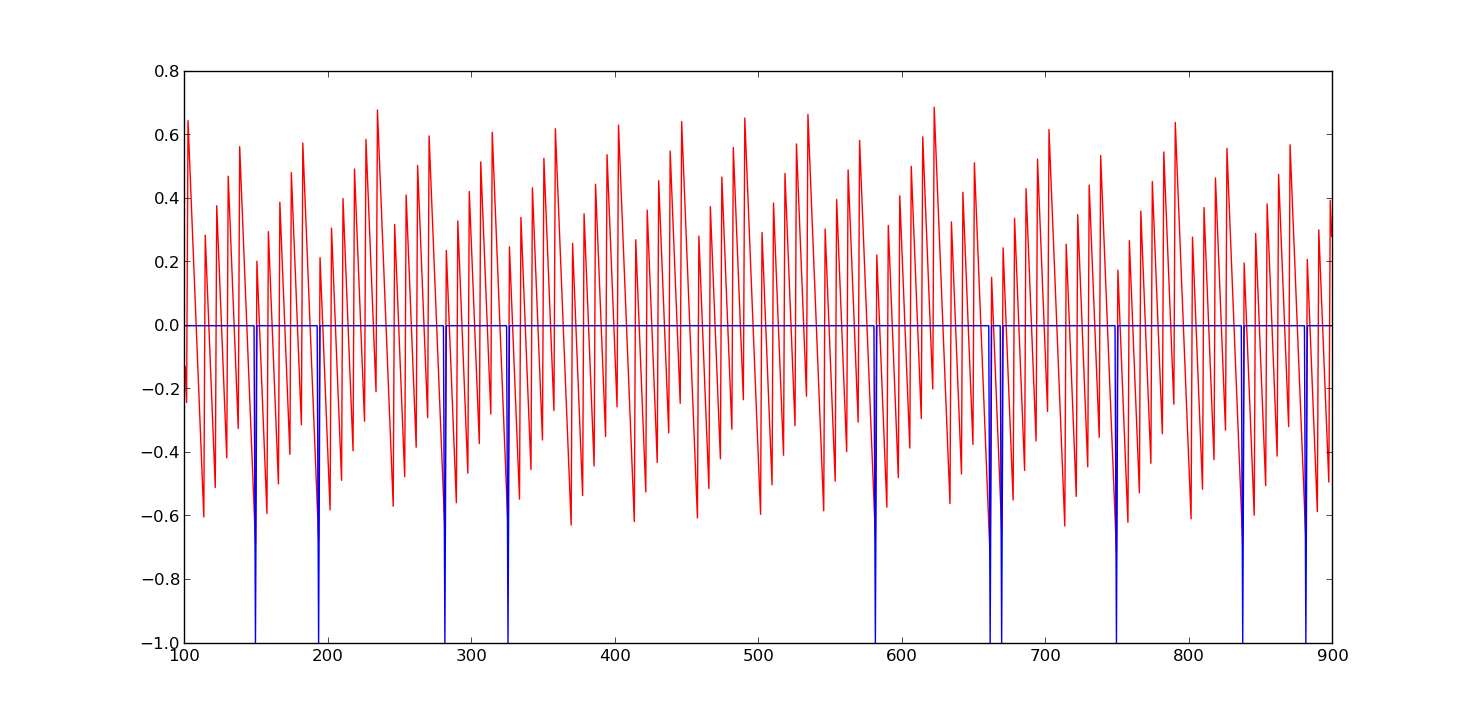}
\caption{For the pattern in Fig. 1, (in red) the fluctuation about the asymptotically expected value $(3/4-2/\pi)N$ of the horizontal extent of the second-farthest vertical line along the positive x-axis and (in blue) the difference between the actual size every fourth step and the nearest integer to $(3/4-2/\pi)(N-3)$}
\label{RvsN}
\end{figure}

We observe that, after an initial transient, $H(N) = \lfloor N/4 \rfloor + $ constant. That is, $H(N)$ increases by 1 after every four steps. Now consider the behaviour of $H'(N)$ with $N$. In fig. \ref{RvsN} (b), we have plotted $H'(N)-(\frac{3}{4}-\frac{2}{\pi})N$ vs. $N$. These fluctuations appear to be bounded, but have a more complicated dependence on N.\\

\begin{figure}[t]
\centering
\includegraphics[scale=0.4]{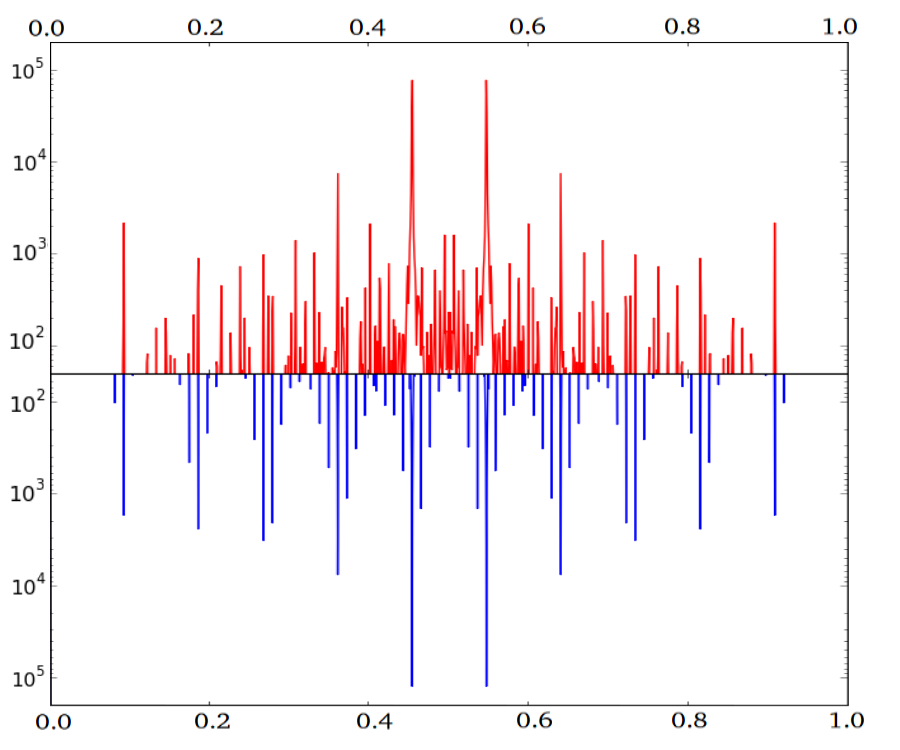}
\caption{The power spectra (on a log scale and arbitrary units) of (in red) the actual sequence $h'_N$ for the pattern and (in blue) the Fourier transform of the ``best approximant" sequence approximating the number $3-\frac{8}{\pi}$}
\label{hvsN}
\end{figure}

To examine these fluctuations in detail, consider the sequence of 1's and 0's in which a `1' denotes that $H'(N)$ increases by 1 during that time-step and `0' denotes that $H'(N)$ stays the same. Denote this sequence by $h'_N$. This also implies that $H'(N) = \sum_{i=0}^N h'_i$. The first thing to note is that $H'(N)$ only increases every fourth step, ie, it never increases when $H(N)$ does not. Thus we can look at $h'_N$ only every fourth step to define a new sequence $h''_N$ for which asymptotic sum approaches the value $\sum_{i=0}^N h''_i = (3-8/\pi)N$. The power spectrum for the sequence $h''_N$ obtained from the pattern is plotted in fig. \ref{hvsN}. There are a number of peaks of various magnitudes, signaling that the sequence might in fact be quasiperiodic.\\

For comparison, consider the fourier transform of the ``best" integer sequence which approximates the irrational growth rate $f = 3-8/\pi$, namely, the sequence $g_N = $ Nint$(f N)$, where Nint$()$ is the nearest-integer function. This Fourier transform is shown in fig. \ref{hvsN} superimposed on the fourier transform of the real sequence. It is seen that the peaks (when the Brillouin zone of frequencies is scaled to go from 0 to 1) for the ``best" sequence are at multiples of $f=3-8/\pi$ to as close an accuracy as the finite length of the sequence allows. Thus, the sequence is quasiperiodic. It is seen that while most of the peaks for the real sequence fall at the same frequencies, their heights are slightly smaller. There are also additional peaks which are not there in the ``best" sequence. To clarify the origin of these peaks, we take a closer look at the difference between the actual sequence and this ``best" one. \\

An equivalent way to form the best sequence is to form the ``best rational approximation" to the number $3-8/\pi$ with denominator at most $N$, and generate a sequence with the periodicity of that rational number. For example, the ``best rational approximation" to the value $1/4$ for any $N>4$ is simply $1/4$, and hence the best such sequence of 1's and 0's which sums to the value $1/4$ for large $N$ is simply 100010001000$\dots$. We calculate the series of ``best rational approximants" to $3-8/\pi$ using ``nearest-integer" continued fractions, which gives us the sequence

\beq
1/2,5/11,39/86,\dots
\eeq

Denote the string `01' by A. For $N<11$, the best sequence would be the sequence AAA$\dots$ such that the total length is less than $N$. For $N>11$, one would need to have five 1's for every seven 0's, and hence the best sequence (for $N<86$) would be $(A^5 0)(A^5 0)(A^5 0)\dots$. We can thus construct these ``best approximant" sequences given an irrational or rational asymptotic growth rate. Note that such a sequence would necessarily by quasiperiodic in the case of an irrational growth rate.\\

Compare this with the actual sequence $h''_N$ from $N/4=242$ to $N/4=511$:

\beqa
&(A^5 0)(A^5 0)(A^5 0)(A^5 0)(A^5 0)(A^5 0)(A^4 0)(A^6 0)(A^4 0)(A^5 0)(A^4 0)(A^6 0)(A^5 0)\\\nonumber
&(A^4 0)(A^6 0)(A^4 0)(A^5 0)(A^5 0)(A^5 0)(A^5 0)(A^5 0)(A^5 0)(A^4 0)(A^6 0)(A^4 0)
\eeqa

It is seen that sometimes the string $(A^5 0)(A^5 0)$, which would give the best approximant, is replaced by the string $(A^4 0)(A^6 0)$. This is probably caused by perturbations due to nearby patches. Thus the sequence $h''_N$, while not the best sequence, still results in bounded fluctuations about the asymptotically expected value.\\

\begin{figure}[t]
\centering
\subfigure[]{
\includegraphics[scale=0.38]{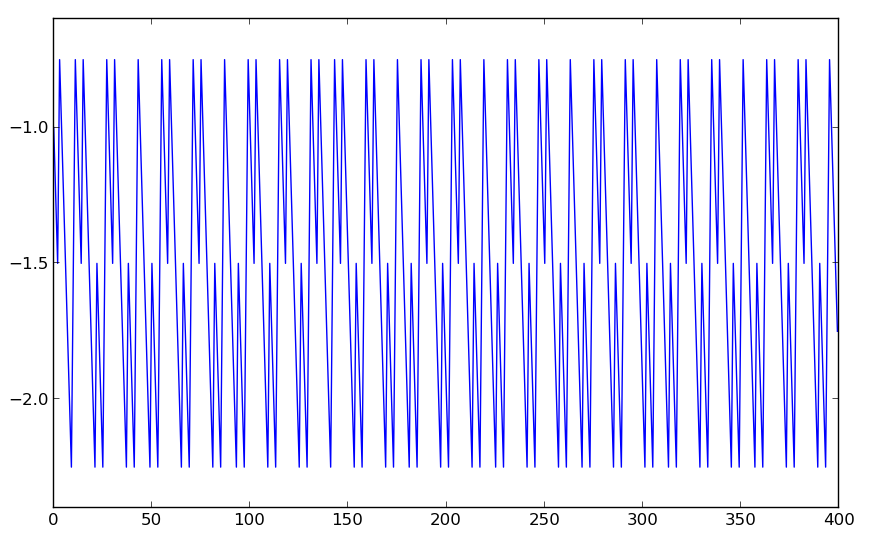}
}
\subfigure[]{
\includegraphics[scale=0.38]{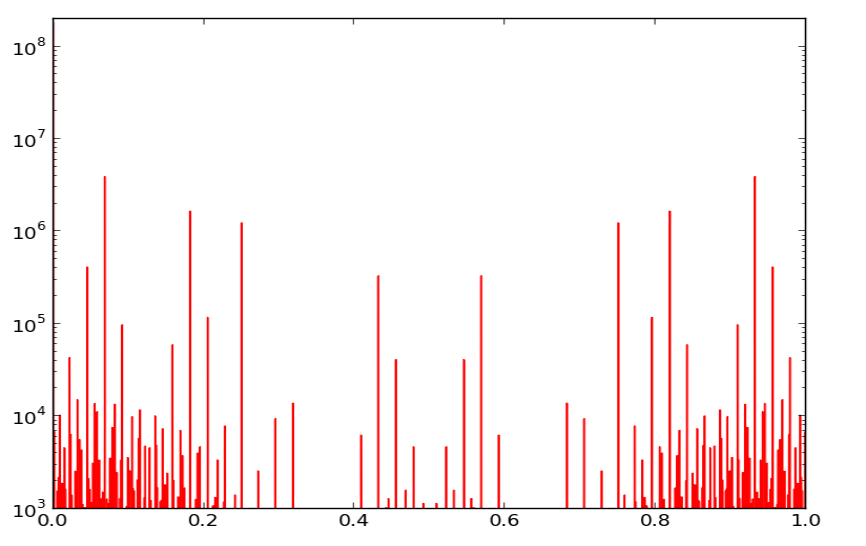}
}
\caption{(a) Fluctuations of the vertical boundary along the positive y-axis, $H_{vert}-N/4$, vs. $N$ and (b) its power spectrum (on a log scale, arbitrary units)}
\label{powerspec}
\end{figure}

This effect of nearby patches is felt even at the boundary of the pattern. Figs. \ref{powerspec} (a) and (b) show $V(N)-N/4$ vs. $N$ and its power spectrum respectively, where $V(N)$ is the vertical extent of the pattern along the positive y-axis. This sequence is not periodic although the associated growth rate is rational, signaling that it is subject to quasiperiodic perturbations from nearby patterns.\\

Our numerical studies suggest that this behaviour, where fluctuations about the asymptotically expected values are bounded and quasiperiodic, is a general property of all macroscopic distances in the pattern.

\section{Summary and concluding remarks}

In this paper, we studied the growing patterns formed by depositing rotor-router walkers on the origin of a finite two-dimensional lattice, starting with a periodic background configurations of arrows. We find that the patterns show rich internal structure which scales proportionately with the size of the pattern. We characterized this structure by mapping it to a resistor network and solving the Kirchoff equations on the network. We also calculated the exact scaling limit  of the  visit function in the asymptotic limit. We found numerical evidence that the fluctuation of the pattern size about the average remain bounded and is a quasiperiodic function of the number of walkers.\\

It would be interesting to explore other  patterns where the basic tiles are not squares. For example, for the sandpile model, patterns on a triangular lattice are found to be hexagonal in shape and composed of triangular tiles \cite{sd12}. The quasiperiodicity and boundedness of the fluctuations in the diameter shown in these rotor-router patterns would be of interest in the more general context of derandomized algorithms.

\appendix

\section{Derivation on eqn. (\ref{eq:fmn}) from matching of boundary conditions}


Fig. \ref{visitpatches} (b) from section \ref{sec:visit} illustrates the arrangement of the six-sided patches of the visit function and its six neighbours, and their co-ordinates on the adjacency graph for the patches. The patch itself has atleast one internal boundary as shown, aligned at 45$^{\circ}$ joining points A and B. The co-ordinates of the points A and B are given by $(f_{m,n-1}-f_{m,n},f_{m-1,n}-f_{m,n})$ and $(f_{m,n+1}-f_{m,n},f_{m+1,n}-f_{m,n})$ respectively. However, since both A and B have to lie on a line of the form $x-y=c$, this implies that the $f$'s must satisfy the equation

\beq
f_{m+1,n} + f_{m-1,n} + f_{m,n-1} + f_{m,n+1} - 4 f_{m,n} = 0
\eeq

This is supplemented with the boundary condition $f(0,0)=0$, since the site $(0,0)$ corresponds to the region outside the octagonal pattern, which has $\phi = 0$.\\

The equation for $f_{m,n}$ is thus the same as that for the infinite square resistor network, and is easily solved to give

\beqa
 f_{m,n} = -\frac{Q}{2\pi^2} \int_{-\pi}^{\pi} \frac{e^{(i k_1 m + k_2 n)}}{1-\frac{1}{2} \sum_i cos(k_i)} \label{eq:a2}
\eeqa

where $Q$ is a constant to be determined. For large $m$ and $n$, that is for $\xi$ and $\eta$ near the origin,

\beqa
\phi \approx m\xi + n\eta - \frac{Q}{2 \pi} \log{(m^2+n^2)}
\eeqa

Note that, as mentioned in section IV, $\nabla^2 \phi_N = N/4$ at the origin, and hence for the the scaled $\phi(\xi,\eta)$ the origin, $\nabla^2 \phi = 2 \delta(\textbf{r})$ (since $D_N = N/2$). Thus, near the origin, $\phi(\xi,\eta)$ behaves like the potential due to a charge with $q=2$ at the origin.

\beqa
 \phi(\textbf{r}) \approx \frac{1}{\pi} \log{|\textbf{r}|}
\eeqa

If a point $\textbf{r}$ near the origin falls in the patch $(m,n)$, $|\textbf{r}| \sim \frac{1}{m^2 + n^2}$. Then equating the function $\phi$ in the above equations for this $\mathbf{r}$, we determine $Q$ in eqn. (\ref{eq:a2}) to be $Q=2$. Thus we have calculated $f_{m,n}$ for the pattern without using the BSST theorem. For example $f_{1,0} = 1/2$ and $f_{2,0} = 1-2/\pi$, giving $R = N/4$ and (as referred to in section \ref{sec:quasi}) $R_1 = N(\frac{3}{4}-\frac{2}{\pi})$.



\end{document}